# $(J + ½)^2$ OR $J(J + 1)$ AS FUNCTIONAL FOR ROTATIONAL TERMS IN ANALYSIS OF MOLECULAR SPECTRA

*J. F. Ogilvie[1]*.

Centre for Experimental and Constructive Mathematics, Department of Mathematics,
Simon Fraser University, 8888 University Drive, Burnaby, British Columbia V5A 1S6 Canada
Escuela de Quimica, Universidad de Costa Rica, Ciudad Universitaria Rodrigo Facio,
San Pedro de Montes de Oca, Costa Rica 11501-2060

**Abstract**

The theoretical and experimental evidence regarding the use of $(J + ½)^2$ and $J(J + 1)$ as a functional in formulae for rotational term values in the spectral analysis of a diatomic molecule in electronic state $^1\Sigma$ is scrutinised. The infrared spectra of HCl serve as examples of the application of the two functionals. The total evidence indicates that $(J + ½)^2$ is preferable to $J(J + 1)$ for the stated purpose, confirming Mulliken's statement in 1930.

**Resumen**

Se analiza la evidencia teórica y experimental sobre el uso de $(J + ½)^2$ y $J(J + 1)$ como un funcional en las fórmulas para los valores de términos rotacionales en el análisis de los espectros de una molécula diatómica en el estado electrónico $^1\Sigma$. El espectro infrarrojo de HCl sirve como ejemplo de la aplicación de los dos funcionales. La total evidencia indica que $(J + ½)^2$ es preferible a $J(J + 1)$ para el uso en el análisis de los espectros, una confirmación de la indicación de Mulliken en 1930.

**Palabras clave:** Espectro rotacional, moléculas diatómicas, términos rotacionales, análisis de espectros
.

## I.  ANALYSIS

Our concern here is with the rotational energy or term values of diatomic molecules, as prototypical of general free molecules, in electronic states of class $^1\Sigma$; the notation for electronic states of other classes can be modified in an appropriate manner.  In accordance with current standard notation in spectral analysis, here we use exclusively symbols $J$ for rotational quantum number and $v$ for vibrational quantum number, regardless of the symbols employed in the original sources.

In his book Molecular Spectra and Molecular Structure, Herzberg (1950) wrote "Some authors prefer to use, instead of

$$F_v(J) = B_v\, J\,(J + 1) - D_v\, J^2\,(J + 1)^2 + \ldots$$

the equation

$$F_v(J) = B_v\,(J + ½)^2 - D_v\,(J + ½)^4 + \ldots$$

which differs from the former equation only by a small additive constant and by a very slight alteration in the meaning of $B_v$ (the new $B_v$ being the old $B_v$ - ½ $D_v$). For the interpretation of the spectra this difference is quite unimportant, but it has to be kept in mind in comparing term values taken from different papers." Herzberg cited no source in this instance, but elsewhere he cited books by Kemble et alii (1927), by Kronig (1930) and by Jevons (1931). The status of the usage of a functional for rotational terms was unsettled during the composition, from 1924 to 1926, of the chapters in the book by Kemble et alii (1927), but the superiority of fitting spectra with $(J + ½)^2$, suggested (Brand, 1995, p.176) by Einstein in 1915, seems to have prevailed. Kronig appears to have employed $J (J + 1)$ in some locations (Kronig, 1930, p. 39) and $(J + ½)^2$ in others (Kronig, 1930, p.31). Although Jevons (1931, p. 24) mentioned $(J + ½)$, he applied $J (J + 1)$ almost exclusively elsewhere. $J (J + 1)$ or $(J + ½)^2$ in those formulae for the rotational term values replaces merely $J^2$ in the preceding quantum theory (Schwarzschild, 1916). Both the latter books (Jevons, 1931; Kronig, 1931) were published after Mulliken's (1930a, p. 65) report in which he stated "It is probable that $(J + ½)^2$ should be used rather than $J (J + 1)$ in [an equation for general term values] and in similar equations, but the difference between the two expressions is negligible for practical purposes, and the form $J(J + 1)$ gives simpler formulas for the frequencies of band lines"; in a subsequent report Mulliken (1930b, p. 621) actually applied $(J + ½)^2$ rather than $J (J + 1)$, consistent with previous experimental work in which he had proved the necessity of half-integer vibrational quantum numbers (Mulliken, 1924). In practice, the formulae for the wavenumbers of lines in a band are no simpler with $J (J + 1)$ than with $(J + ½)^2$ as functional, cf. the appendix; with contemporary methods of fitting spectra any such distinction is of negligible concern.

---
[1] Corresponding author: ogilvie@cecm.sfu.ca

As an instance in which ½ appears in a formula for rotational terms we recall the report of Colby (1923) on the analysis of the absorption spectrum of HCl in the mid- and near-infrared regions, in which he stated this formula for the energies of rotational states,

$$W_{vJ} = W_{v0} + h\, B_v\, (J + ½)^2 - h\, \beta\, (J + ½)^4$$

involving Planck constant $h$ and parameters $B_v$ and $\beta$ that have frequency units. In the conclusion of this paper (Colby, 1923), the author stated "On the whole one may feel that this is satisfactory evidence of the necessity for introducing half parameter numbers into the formulations of these bands".

There is no question that, for the hydrogen atom, the squared angular momentum should be based on $l (l + 1)$, as Schroedinger (1926a) obtained, but a diatomic molecule is an entity with properties distinct from those of a hydrogen atom, even if there be a formal correspondence between the Coulomb problem and a quadratic harmonic oscillator in multiple dimensions (Kostelecky et alii, 1985). In the second paper in the series Quantisation as a Problem of Proper Values, Schroedinger (1926b) solved his equation for 1) a 'Planck oscillator' – i.e. a canonical linear harmonic oscillator with a quadratic dependence of the potential energy on the displacement from an equilibrium condition, 2) a rotor with a fixed axis that produced an energy proportional to the square of a quantum number, 3) a rigid rotor with a free axis that produced a rotational energy based on functional $J (J + 1)$ and 4) a non-rigid rotor as a model of a diatomic molecule that produced again an energy dependent on the same functional. Elsewhere, Schroedinger (1926e) subsequently wrote "It is well known that so-called half-quantum numbers are actually supported by the experimental evidence on most of the simple band spectra, and are probably contradicted by none of them. Mr. Fues … has worked out the band theory of diatomic molecules in detail, taking into account the mutual influence of rotation and oscillation and the fact that the latter is not of the simple harmonic type. The result is in exact agreement with the ordinary treatment except that the quantum numbers become half-integer also in all correction terms." As a basis of his solution of Schroedinger's equation for an oscillator, Fues (1926) applied the potential-energy function of Kratzer (1920), which is expressed in simplified form as

$$V(R) = h\, c\, \mathcal{D}_e\, (1 - R_e/R)^2$$

Therein appear speed $c$ of light in vacuo, $R$ that would represent the instantaneous internuclear separation of a diatomic molecule, equilibrium internuclear distance $R_e$ and energy $h\, c\, \mathcal{D}_e$ at the dissociation limit relative to the energy for $R = R_e$. The results of those calculations yielded an expression for vibration-rotational terms that includes $(J + ½)$ to even powers (Fues, 1926; Kronig, 1931).

The quadratic linear harmonic oscillator that Schroedinger (1926b) treated is a poor model for a real diatomic molecule because, in addition to states of infinite number and thus no limit that would correspond to a molecular dissociation into atomic fragments, rotational parameter $B_v$ for a particular vibrational state increases appreciably – at least 3 per cent – with each increment of vibrational quantum number $v$, contrary to the decrease that is found for any real diatomic molecule; for instance, for HCl the decrease is at least 3 per cent. A superior model function is the oscillator of Davidson (1932), of which the potential energy has this form in terms of spectral parameters $\omega_e$ and $B_e$:

$$V(R) = \omega_e^2 \, (R/R_e - R_e/R)^2 \, / 16 \, B_e$$

Although this function still has equal intervals between states of adjacent energy such that it qualifies directly as an harmonic oscillator, and has hence states of infinite number, its rotational parameters are constant, independent of $v$ – i.e. this harmonic oscillator is effectively also a rigid rotor. The most notable property of this oscillator is that it automatically generates a contribution ¼ $B_e$ to the energy of each state characterised with vibrational quantum number $v$. Another property of significant interest is that matrix elements of $R$, to the square of which the intensities of vibrational transitions are proportional, are finite for $\Delta v > 1$; the magnitudes of those matrix elements for transitions from $v = 0$ are comparable with those of HCl for its overtone bands. Davidson's (1932) oscillator is clearly a model preferable to that of the quadratic oscillator for practical purposes.

The ultimate treatment of a rotating oscillator to serve as a quantitative model of a diatomic molecule is due to Dunham (1932b) who applied a general polynomial function for vibrational potential energy involving coefficients $a_j$ with reduced displacement variable $x = (R - R_e)/R_e$:

$$V(x) = a_0 \, x^2 \, (1 + \sum_{j=1} a_j \, x^j)$$

Dunham (1932b) expressed the vibration-rotational terms, which he obtained with a JBKW method (Dunham, 1932a) but which are equally well and more conveniently derived with hypervirial perturbation theory (Fernandez and Ogilvie, 1990), as double sums over $(v + ½)$ and $J(J + 1)$, the latter being an assumed functional form not explicitly justified:

$$E_{v,J} = \sum_{k=0} \sum_{l=0} Y_{k,l} \, (v + ½)^k \, [J(J + 1)]^l$$

Coefficient $Y_{0,0}$, which, with ½ $Y_{1,0}$, is obviously a contribution to the residual (or 'zero-point') energy, has, apart from contributions beyond least order, exactly the following form as expressed (Dunham, 1932b) in terms of coefficients $a_1$ and $a_2$ in the above function for potential energy,

$$Y_{0,0} = B_e \, (-7 \, a_1^2 \, / 32 \, + \, 3 \, a_2 \, / \, 8)$$

but is expressible alternatively (Bunker, 1968) in terms of traditional spectral parameters as

$$Y_{0,0} \approx ¼ \, B_e + \alpha_e \, \omega_e \, /12 \, B_e + \alpha_e^2 \, \omega_e^2 \, /144 \, B_e^3 \, - \, ¼ \, \omega_e x_e$$

The first term of this formula is hence precisely the difference between the leading terms in the two formulae for rotational terms quoted from Herzberg's (1950) book above; if that quantity ¼ $B_e$ be transferred from $Y_{0,0}$ to $Y_{0,1}$ so that the double sum contains $(J + ½)^2$ instead of $J(J + 1)$, the three remaining quantities in $Y_{0,0}$ tend to cancel one another (Ogilvie, 1987). After such a transfer to modify the nature of Dunham's double sum, the nature of $Y_{0,1}$ would alter slightly (Herzberg, 1950), and the numerical values of other fitted coefficients $Y_{k,l}$ with $l > 0$ would also be only slightly altered, but not their physical significance; the modified definition of coefficient $Y_{0,0}$ to second order would become

$$Y_{0,0} = B_e \, (-7 \, a_1^2 \, /32 + 3 \, a_2 \, /8 - ¼ \, )$$

From an experimental point of view, for any spectra involving changes of rotational quantum number $J$ between combining states, the presence or absence of ¼ $B_e$ in the functional is immaterial because this quantity cancels in the difference between the rotational terms of those states. Although one might derive directly $J(J + 1)$ from such an empirical analysis of rotational spectra (Monagan and Ogilvie, 1987), in fact $J^2 + J + n$ is compatible with such an analysis, with $n$ taking an arbitrary numerical value; the only two values of $n$ worth considering are 0 and ¼. One might attempt to measure some transition of isotopically related molecules to a state involving no rotational angular momentum. The threshold energies, expressed as spectral terms, to dissociate, for instance, HCl and DCl, into their atomic fragments have been accurately measured (Martin and Hepburn, 1998; Hu et alii, 2003) as follows.

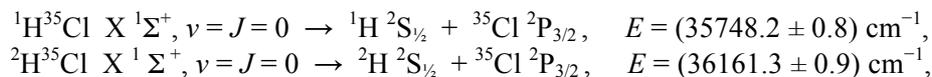

$^1H^{35}Cl\ X\ ^1\Sigma^+, v = J = 0 \rightarrow\ ^1H\ ^2S_{1/2} + ^{35}Cl\ ^2P_{3/2},\quad E = (35748.2 \pm 0.8)\ cm^{-1},$
$^2H^{35}Cl\ X\ ^1\Sigma^+, v = J = 0 \rightarrow\ ^2H\ ^2S_{1/2} + ^{35}Cl\ ^2P_{3/2},\quad E = (36161.3 \pm 0.9)\ cm^{-1},$

The difference, $(413.1 \pm 1.2)\ cm^{-1}$, between these two values is expected to correspond to the difference of the total residual energies, which would hence consist of both vibrational and rotational contributions if the latter exist (Ogilvie, 2014). The sum with evaluated Dunham coefficients (Uehara et alii, 2004) involving only the vibrational terms,

$$G(v) = \sum_{k=0}^{5} Y_{k,0}\ (v + \tfrac{1}{2})^k$$

for $v = 0$ yields $(1483.88 \pm 0.1)\ cm^{-1}$ for $^1H^{35}Cl$ and $(1066.61 \pm 0.1)\ cm^{-1}$ for $^2H^{35}Cl$. Their difference, $(417.3 \pm 0.1)\ cm^{-1}$, is outside $(3.5\ \sigma)$ the experimental uncertainty of the difference of the measured terms for dissociation into neutral atoms, and likewise outside the uncertainty of the difference, $(442.9 \pm 0.9)\ cm^{-1}$, of the measured terms for dissociation into ions with account taken of the difference, $(29.8 \pm 0.1)\ cm^{-1}$, between the ionisation terms for $^1H$ and $^2H$ (Ogilvie, 2014; Uehara et alii, 2004).

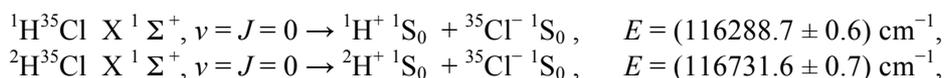

$^1H^{35}Cl\ X\ ^1\Sigma^+, v = J = 0 \rightarrow\ ^1H^+\ ^1S_0 + ^{35}Cl^-\ ^1S_0,\quad E = (116288.7 \pm 0.6)\ cm^{-1},$
$^2H^{35}Cl\ X\ ^1\Sigma^+, v = J = 0 \rightarrow\ ^2H^+\ ^1S_0 + ^{35}Cl^-\ ^1S_0,\quad E = (116731.6 \pm 0.7)\ cm^{-1},$

From these ionisation terms, the preceding dissociation terms were derived (Martin and Hepburn, 1998; Hu et alii, 2003). There is hence no satisfactory quantitative agreement, within the experimental uncertainties, between the calculated residual energies and the difference between the measured dissociation energies; this result is independent of whether quantity $\tfrac{1}{4} B_e$ be included either within $Y_{0,0}$ or taken into account with functional $(J + \tfrac{1}{2})^2$. Other and similar comparisons (Hu et alii, 2003) are likewise inconclusive. At this time there seems to exist no definitive experimental test of the existence of such a residual rotational energy in a distinguishable form.

Another aspect of a consideration of rotational and vibrational terms is their combined treatment according to fractional calculus (Hermann, 2013, 2014) to overcome the traditional distinction between vibrational and rotational degrees of freedom. For this purpose the vibrational, $(v + \tfrac{1}{2})$, and rotational, $(J + \tfrac{1}{2})^2$, functionals must have the forms as explicitly specified. Fractional parameter $\alpha$ as $(n + \tfrac{1}{2})^\alpha$ with $n = v$ or $J$ allows for a smooth transition between the two limiting cases, $\alpha = 1$ and $\alpha = 2$. An application, to only third order in anharmonic corrections and centrifugal stretching, of this approach to the absorption spectrum of gaseous HCl in the mid-infrared region yielded a calculated spectrum (Hermann, 2013) of which both the wavenumbers and the intensities of the lines approximate the experimental spectrum moderately accurately.

In summary, the bulk of the available theoretical evidence indicates that, as a functional in the formula for rotational term values, $(J + \tfrac{1}{2})^2$ is preferable to $J(J + 1)$; there appears to be no experimental evidence to the contrary. Mulliken (1930a,b) recognised correctly that $(J + \tfrac{1}{2})^2$ should be used rather than $J(J + 1)$ in the formula for rotational term values, but he erred in proceeding to endorse the latter for use in the analysis of molecular spectra involving rotational contributions to the transitions. Future spectral analyses should hence be undertaken with the correct functional, $(J + \tfrac{1}{2})^2$, but any use of either functional in an application must be stated explicitly.

## APPENDIX

We present here several formulae, not published elsewhere, that are directly useful in an analysis of pure-rotational and vibration-rotational spectra, in both absorption or emission and Raman scattering, in terms of (J + ½) to various powers for the rotational terms. Explicitly, for the pure rotational spectrum we express these terms, implicitly in wavenumber unit, as

$$F(J) = B\,(J+\tfrac{1}{2})^2 - D\,(J+\tfrac{1}{2})^4 + H\,(J+\tfrac{1}{2})^6 + L\,(J+\tfrac{1}{2})^8$$

involving conventional parameters $B$, $D$, $H$ and $L$, and retaining contributions up to $J^8$. Accordingly, for lines in a pure-rotational spectrum in absorption or emission within one vibrational state with selection rule $\Delta J = +1$, implying branch R, the wavenumbers of the lines conform to this formula.

$$R(J) = \left(2B - D + \frac{3H}{8} + \frac{L}{8}\right)(J+1) + \left(-4D + 5H + \frac{7L}{2}\right)(J+1)^3$$
$$+ (6H + 14L)(J+1)^5 + 8L(J+1)^7$$

For purposes of linear regression to evaluate spectral coefficients $B$, $D$,..., each with its associated standard deviation, this form is practical.

$$R(J) = (2J+2)B + \left(-J-1-4(J+1)^3\right)D + \left(\frac{3J}{8} + \frac{3}{8} + 5(J+1)^3 + 6(J+1)^5\right)H$$
$$+ \left(\frac{J}{8} + \frac{1}{8} + \frac{7(J+1)^3}{2} + 14(J+1)^5 + 8(J+1)^7\right)L$$

For lines in the pure-rotational spectrum within one vibratioinal state recorded as Raman scattering according to selection rule $\Delta J = +2$, implying branch S, the wavenumbers of the lines conform to this formula.

$$S(J) = (4B - 8D + 12H + 16L)\left(J+\tfrac{3}{2}\right) + (-8D + 40H + 112L)\left(J+\tfrac{3}{2}\right)^3$$
$$+ (12H + 112L)\left(J+\tfrac{3}{2}\right)^5 + 16L\left(J+\tfrac{3}{2}\right)^7$$

For the same purpose of separate evaluation of each spectral coefficient, the following formula is applicable.

$$S(J) = (4J+6)B + \left(-8J - 12 - 8\left(J+\tfrac{3}{2}\right)^3\right)D$$
$$+ \left(12J + 18 + 40\left(J+\tfrac{3}{2}\right)^3 + 12\left(J+\tfrac{3}{2}\right)^5\right)H$$
$$+ \left(16J + 24 + 112\left(J+\tfrac{3}{2}\right)^3 + 112\left(J+\tfrac{3}{2}\right)^5 + 16\left(J+\tfrac{3}{2}\right)^7\right)L$$

For vibration-rotational spectra that involve transitions between rotational terms belonging to two vibrational states, state 2 having a vibrational term greater than state 1 by $\nu_0$, the formula for the rotational terms within each state is the same as above, but we must distinguish the spectral coefficients of each state. For electronic state of class $^1\Sigma$ there occur two branches of a band that comprises the absorption or emission spectrum, branch R with $\Delta J = +1$,

$$R(J) = \nu_0 + F(J+1, B_2, D_2, H_2, L_2) - F(J, B_1, D_1, H_1, L_1)$$

and branch P with $\Delta J = -1$,

**$(J + ½)^2$ OR $J(J + 1)$ AS FUNCTIONAL FOR ROTATIONAL TERMS IN ANALYSIS OF MOLECULAR SPECTRA**

$$P(J) = \nu_0 + F(J - 1, B_2, D_2, H_2, L_2) - F(J, B_1, D_1, H_1, L_1)$$

An elegant method of evaluating separately the spectral coefficients of each vibrational state involves the use of combination differences. This combination,

$$R(J) - P(J) = (4 B_2 - 8 D_2 + 12 H_2 + 16 L_2)\left(J + \frac{1}{2}\right)$$
$$+ (-8 D_2 + 40 H_2 + 112 L_2)\left(J + \frac{1}{2}\right)^3 + (12 H_2 + 112 L_2)\left(J + \frac{1}{2}\right)^5 + 16 L_2 \left(J + \frac{1}{2}\right)^7$$

yields the parameters of state 2, whereas this combination,

$$R(J - 1) - P(J + 1) = (4 B_1 - 8 D_1 + 12 H_1 + 16 L_1)\left(J + \frac{1}{2}\right)$$
$$+ (-8 D_1 + 40 H_1 + 112 L_1)\left(J + \frac{1}{2}\right)^3 + (12 H_1 + 112 L_1)\left(J + \frac{1}{2}\right)^5 + 16 L_1 \left(J + \frac{1}{2}\right)^7$$

yields the parameters of state 1. For the purpose of linear regression that yields directly the values and standard deviations of the separate spectral parameters, these two formulae are applicable.

$$R(J) - P(J) = (4 J + 2) B_2 + \left(-8 J - 4 - 8\left(J + \frac{1}{2}\right)^3\right) D_2$$
$$+ \left(12 J + 6 + 40\left(J + \frac{1}{2}\right)^3 + 12\left(J + \frac{1}{2}\right)^5\right) H_2$$
$$+ \left(16 J + 8 + 112\left(J + \frac{1}{2}\right)^3 + 112\left(J + \frac{1}{2}\right)^5 + 16\left(J + \frac{1}{2}\right)^7\right) L_2$$

$$R(J - 1) - P(J + 1) = (4 J + 2) B_1 + \left(-8 J - 4 - 8\left(J + \frac{1}{2}\right)^3\right) D_1$$
$$+ \left(12 J + 6 + 40\left(J + \frac{1}{2}\right)^3 + 12\left(J + \frac{1}{2}\right)^5\right) H_1$$
$$+ \left(16 J + 8 + 112\left(J + \frac{1}{2}\right)^3 + 112\left(J + \frac{1}{2}\right)^5 + 16\left(J + \frac{1}{2}\right)^7\right) L_1$$

The following combination sum enables the evaluation of vibrational term difference $\nu_0$ corresponding to the band origin, after the values of the pertinent rotational parameters have been evaluated with the above formulae.

$$R(J - 1) + P(J) = \left(2 \nu_0 + \frac{1}{2} B_2 - \frac{1}{8} D_2 + \frac{1}{32} H_2 + \frac{1}{128} L_2 - \frac{1}{2} B_1 + \frac{1}{8} D_1 - \frac{1}{32} H_1 - \frac{1}{128} L_1\right)$$
$$+ \left(2 B_2 - 3 D_2 + \frac{15}{8} H_2 + \frac{7}{8} L_2 - 2 B_1 + 3 D_1 - \frac{15}{8} H_1 - \frac{7}{8} L_1\right) J^2 +$$
$$\left(-2 D_2 + \frac{15}{2} H_2 + \frac{35}{4} L_2 + 2 D_1 - \frac{15}{2} H_1 - \frac{35}{4} L_1\right) J^4 + (2 H_2 + 14 L_2 - 2 H_1 - 14 L_1) J^6 + (2 L_2 - 2 L_1) J^8$$

In a vibration-rotational band recorded as Raman scattering, there occur two branches S, for $\Delta J = +2$, and O, $\Delta J = -2$, on one or other side of central branch Q, for $\Delta J = 0$.

$$S(J) = \nu_0 + F(J+2, B_2, D_2, H_2, L_2) - F(J, B_1, D_1, H_1, L_1)$$

$$O(J) = \nu_0 + F(J-2, B_2, D_2, H_2, L_2) - F(J, B_1, D_1, H_1, L_1)$$

The combination differences appropriate to the separate evaluation of the rotational parameters of each vibrational state conform to these formulae.

$$S(J) - O(J) = (8\,B_2 - 64\,D_2 + 384\,H_2 + 2048\,L_2)\left(J+\frac{1}{2}\right)$$
$$+ (-16\,D_2 + 320\,H_2 + 3584\,L_2)\left(J+\frac{1}{2}\right)^3 + (24\,H_2 + 896\,L_2)\left(J+\frac{1}{2}\right)^5 + 32\,L_2\left(J+\frac{1}{2}\right)^7$$

$$S(J-2) - O(J+2) = (8\,B_1 - 64\,D_1 + 384\,H_1 + 2048\,L_1)\left(J+\frac{1}{2}\right)$$
$$+ (-16\,D_1 + 320\,H_1 + 3584\,L_1)\left(J+\frac{1}{2}\right)^3 + (24\,H_1 + 896\,L_1)\left(J+\frac{1}{2}\right)^5 + 32\,L_1\left(J+\frac{1}{2}\right)^7$$

The corresponding formulae to evaluate the rotational parameters pertaining to the separate vibrational states follow.

$$S(J) - O(J) = (8J+4)\,B_2 + \left(-64J - 32 - 16\left(J+\frac{1}{2}\right)^3\right)D_2$$
$$+ \left(384J + 192 + 320\left(J+\frac{1}{2}\right)^3 + 24\left(J+\frac{1}{2}\right)^5\right)H_2$$
$$+ \left(2048J + 1024 + 3584\left(J+\frac{1}{2}\right)^3 + 896\left(J+\frac{1}{2}\right)^5 + 32\left(J+\frac{1}{2}\right)^7\right)L_2$$

$$S(J-2) - O(J+2) = (8J+4)\,B_1 + \left(-64J - 32 - 16\left(J+\frac{1}{2}\right)^3\right)D_1$$
$$+ \left(384J + 192 + 320\left(J+\frac{1}{2}\right)^3 + 24\left(J+\frac{1}{2}\right)^5\right)H_1$$
$$+ \left(2048J + 1024 + 3584\left(J+\frac{1}{2}\right)^3 + 896\left(J+\frac{1}{2}\right)^5 + 32\left(J+\frac{1}{2}\right)^7\right)L_1$$

After the evaluation of the separate rotational parameters, the combination sum to evaluate the band origin, $\nu_0$, has this formula.

$$S(J+2) + O(J) =$$
$$(2\nu_0 + 18\,B_2 - 162\,D_2 + 1458\,H_2 + 13122\,L_2 - 2\,B_1 + 2\,D_1 - 2\,H_1 - 2\,L_1) +$$
$$(2\,B_2 - 108\,D_2 + 2430\,H_2 + 40824\,L_2 - 2\,B_1 + 12\,D_1 - 30\,H_1 - 56\,L_1)\left(J+\frac{3}{2}\right)^2 +$$
$$(-2\,D_2 + 270\,H_2 + 11340\,L_2 + 2\,D_1 - 30\,H_1 - 140\,L_1)\left(J+\frac{3}{2}\right)^4 +$$
$$(2\,H_2 + 504\,L_2 - 2\,H_1 - 56\,L_1)\left(J+\frac{3}{2}\right)^6 + (2\,L_2 - 2\,L_1)\left(J+\frac{3}{2}\right)^8$$

These formulae have direct practical application to evaluate the pertinent spectral parameters of a pure-rotational or vibration-rotational band in conformity with a rotational term value expressed in terms of (*J* + ½) to various powers. If not all contributions to the rotational terms up to $(J + ½)^8$ are necessary, each formula can be truncated at an appropriate level.